# THE EFFECT OF TREATMENT-RELATED DEATHS AND "STICKY" DIAGNOSES ON RECORDED PROSTATE CANCER MORTALITY

H. Gilbert Welch, MD, MPH

Michael J. Barry MD

William C Black MD

Yunjie Song, PhD

Elliott S. Fisher MD, MPH

From the VA Outcomes Group, Department of Veterans Affairs Medical Center, White River Junction, VT (HGW, ESF) the Dartmouth Institute for Health Policy and Clinical Practice, Dartmouth Medical School, Hanover, NH (HGW, WCB, YS, ESF) and the Massachusetts General Hospital, Harvard Medical School, Boston, MA (MJB).

This study was supported by a R01 from the National Cancer Institute (CA107124). The views expressed herein do not necessarily represent the views of the Department of Veterans Affairs or the United States Government.

Author's Note:
This manuscript was written in 2008. It was reviewed (with revisions requested) and resubmitted at both the New England Journal and the Journal of the National Cancer Institute – but was ultimately rejected by each. *[The first author bears sole responsibility for not further pursuing its publication.]* It is being posted a decade later as it may have relevance to understanding the rise and fall in prostate cancer mortality following the growth of TURP (transurethral resection of the prostate) and the advent of PSA (prostate specific antigen) screening.

H. Gilbert Welch [drgilwelch@gmail.com]
Senior Investigator
The Center for Surgery and Public Health
Brigham and Women's Hospital, Boston

**ABSTRACT**

**Background**: Although recorded cancer mortality should include both deaths from cancer and deaths from cancer treatment, there is some evidence suggesting that the measure may be incomplete. To investigate the completeness of recorded prostate cancer mortality, we compared other-cause (non-prostate cancer) mortality in men found and not found to have prostate cancer following a needle biopsy.

**Methods**: We linked Medicare claims data to Surveillance, Epidemiology, and End Results (SEER) data to analyze survival in the population of men aged 65 years and older enrolled in Medicare Part B who resided in a SEER area and who received a needle biopsy of the prostate in 1993 through 2001. We compared other-cause mortality in men found to have prostate cancer (n= 53,462) to that in men not found to have prostate cancer (n = 103,659).

**Results**: The age-race adjusted other-cause mortality rate was 471 per 10,000 person-years in men found to have prostate cancer vs. 468 per 10,000 in men not found to have prostate cancer (RR = 1.01; 95% CI: 0.98 - 1.03). The effect was modified, however, by age. The RR declined in a stepwise fashion from 1.08 (95% CI 1.03 – 1.14) in men age 65-69 to 0.89 (95% CI 0.83 – 0.95) in men age 85 and older. If the excess (or deficit) in other-cause mortality were added to the recorded prostate cancer mortality, prostate cancer mortality would rise as much as 23% in the youngest age group (from 90 to 111 per 10,000) and would fall as much as 30% in the oldest age group (from 551 to 388 per 10,000).

**Conclusion**: Although recorded prostate cancer mortality appears to be an accurate measure overall, it systematically underestimates the mortality associated with prostate cancer diagnosis and treatment in younger men and overestimates it in the very old. We surmise that in younger men treatment-related deaths are incompletely captured in recorded prostate cancer mortality, while in older men the diagnosis "sticks" – once diagnosed, they are more likely to be said to have died from the disease.

The best indicator of progress against cancer is a decline in population-based cancer mortality rates.[1] One strategy to achieve such progress is early diagnosis. But because early diagnosis almost always means more patients are diagnosed and treated for cancer than would be otherwise, it is possible that while mortality from advanced cancer falls – treatment-related mortality rises. Thus to ensure that progress is genuine, cancer mortality should include not only deaths from cancer, but also deaths from cancer treatment.

There are reasons to worry that treatment-related mortality may not be reliably included in the measurement of cancer mortality. Although the World Health Organization's definition of the underlying cause of death clearly encompasses treatment-related mortality ("the disease or injury which initiated the train of morbid events leading directly to death")[2], the instructions of US National Center for Heath Statistics on coding the underlying cause of death contain conflicting guidance on whether treatment-related mortality should be attributed to cancer or to an "accident" (i.e. a medical misadventure)[3]. Empirical evidence bears out that some confusion exists. One of us (HGW) reported that 41% of deaths within a month following cancer-directed surgery were not attributed to the cancer for which the surgery was performed.[4] And Brown et al. reported that other-cause (non-cancer) mortality was considerably higher in cancer patients than in the general population - particularly in the year immediately following the diagnosis.[5]

Enumerating treatment-related mortality may be particularly important in prostate cancer given the increasing use of androgen deprivation therapy, not only for metastatic disease, but also for earlier stage disease. Roughly a quarter-million Medicare eligible men now receive this therapy – almost 3% of all men in Medicare[6]. A recent meta-analysis of (conducted as part of an American Society of Clinical Oncology practice guideline update) suggested that androgen deprivation therapy may be associated with a 15% increase in other-cause mortality[7], perhaps reflecting an increased risk of cardiovascular disease[8]. These deaths would be unlikely to be attributed to prostate cancer in vital statistics.

In this paper, we refine Brown's approach to consider the validity of recorded prostate cancer mortality.  Here there is little doubt about the effect of early diagnosis on the number of men treated for the disease - hundreds of thousands of additional men have been treated for prostate cancer following the introduction of the PSA.  Thus even if the individuals mortality risk from treatment were small, recorded prostate cancer mortality could be substantially undercounted were treatment-related mortality not included.  We hypothesized that among a cohort of men undergoing prostate biopsy this undercount would appear as elevated other-cause mortality in those found to have prostate cancer relative to those <u>not</u> found to have prostate cancer.

**METHODS**

*Overview*

To make an inference about the completeness of recorded prostate cancer mortality, we compared other-cause mortality in men found and not found to have prostate cancer following a needle biopsy. We postulated that by starting with a cohort of men undergoing biopsy we could construct groups that were otherwise comparable - with the exception of the known prostate cancer risk factors of age and race. In other words, although some men would be found to have prostate cancer and some would not, the two groups would be otherwise similar because they share the same characteristics that led to biopsy. Thus if age-race adjusted other-cause mortality was higher in the group diagnosed with prostate cancer, we would infer that measured prostate cancer mortality was incomplete.

Mechanically, the analysis required two data sources. We used the Medicare National Claims History System (Source: Health Care Financing Agency; Baltimore, MD) both to identify the cohort of men undergoing biopsy and to determine the fact of death. We used the Surveillance Epidemiology and End Results (SEER) data to determine which men were diagnosed with prostate cancer and to determine their cause of death.

Although a linked Medicare-SEER dataset already exists, it only includes data on beneficiaries with cancer and a 5% sample of controls. Because our analysis required survival data for an entire population of men undergoing prostate biopsy (both those who did and did not receive a prostate cancer diagnosis), we needed to create a new Medicare-SEER linkage. This study was approved by the Dartmouth institutional review board.

*Study Population*

We identified all men with a Medicare Part B claim for a needle biopsy of the prostate (identified using the Current Procedure Terminology (CPT) code 55700) who resided in a SEER area during the period 1993-2001. We first used the Medicare denominator file to restrict the analysis to men who resided in one of the SEER 11 areas (5 states: Connecticut, Hawaii, Iowa, New Mexico, and Utah; 6 metropolitan areas Atlanta, Detroit, Los Angeles, San Francisco, San Jose, and Seattle) both in the year of the biopsy and the subsequent year. In other words, we created a dataset of prostate biopsies performed in men who we could ensure resided in a SEER area at the time of the biopsy – a restriction necessary to capture the diagnosis of prostate cancer in SEER.

We then excluded three groups using demographic exclusion criteria. First, we excluded biopsies in men who were not age 65 and older at the start of each calendar year (both because individuals who become eligible for Medicare for reasons other than age represent a very unique group and because those who become eligible mid-year will not have complete data). Second, we excluded those men not entitled to Part B for the entire year (as diagnostic services provided to them are not reported in the claims). Third, we excluded those in men enrolled in risk-contract managed care plans (for the same reason).

We then excluded two more groups of biopsies using clinical exclusion criteria. First, we excluded biopsies in men with a concomitant TURP (transurethral resection of the prostate), defined as one within a five month window surrounding the prostate biopsy, as it would not be clear whether to attribute a prostate cancer diagnosis to the TURP or the biopsy. Second, we excluded biopsies in men with a prior prostate cancer diagnosis. Following these exclusions our study cohort consisted of 157,121 men undergoing needle biopsy of the prostate.

*Exposure: Prostate Cancer Diagnosis*

To determine which of these men were diagnosed with prostate cancer, we linked the Medicare beneficiary ID to the SEER data. A member of our cohort was determined to develop prostate cancer if he appeared in the SEER data as an incident cancer case with a primary site International Classification of Disease – Oncology (ICD-O) code of C619.

As in our prior work[9], we considered a prostate cancer diagnosis within 1 month prior to and 3 months after the biopsy date as being attributable to the biopsy (SEER only codes the month of diagnosis - given a biopsy date of 12/4/93, a prostate cancer diagnosis accessed in 11/93, 12/93 or 1/94, 2/94, 3/94 would be attributed to the biopsy). Our use of this 5 month window reflected the empirical observations that an excess number of prostate cancer diagnoses appeared in this period: some in the month prior to biopsy (likely representing SEER linking the date of diagnosis to the PSA test prompting the biopsy) and some in the months following the biopsy (representing a time lag). Using our SEER-Medicare linkage we determined that 53,462 men were found to have prostate cancer following the needle biopsy and 103,659 men were not found to have prostate cancer.

*Outcome: Other Cause Mortality*

Our primary outcome is other-cause mortality (expressed as deaths per 10,000 person years). For all members in our cohort, the fact of death and its date was obtained from Medicare denominator file. For the 53,462 men found to have prostate cancer (who appear in the SEER data), the underlying cause of death was obtained in SEER data. Other-cause mortality was determined by removing those deaths attributed to prostate cancer (those that encompass measured prostate cancer mortality) from all deaths combined. Because the 103,659 men <u>not</u> found to have prostate cancer either never receive the diagnosis of prostate cancer or are censored when they do (as detailed below), other-cause mortality simply reflects all deaths combined.

*Survival Analysis*

Given that members of our cohort could enter any time over the 9-year period (1993-2001), the length of follow-up was variable. To deal with the problem of variable follow-up we used survival analysis.

For all members of the cohort, survival time is calculated from the date of the index biopsy. The failure event is a death from something other than prostate cancer (date = date of death). The most frequent censoring event was simply the study closure date – chosen to provide a minimum of 2 years of follow-up (date = 12/31/03). Members were also censored at the time relevant data were no longer available: moving out of a SEER area and leaving Medicare Part B (either because of not purchasing it or because of entering a risk-contract HMO).

Two censoring events were distinct among the two groups. Men found to have prostate cancer were censored if they died from prostate cancer. In other words, on the date of prostate cancer death they are no longer "at risk" for the failure event - death from something other than prostate cancer. Men not found to have prostate cancer were censored if they were diagnosed with prostate cancer. At this point, they are no longer a member of their original exposure group. **Figure 1** shows the study overview and the number experiencing the various censoring events.

**Figure 1: Study Overview**

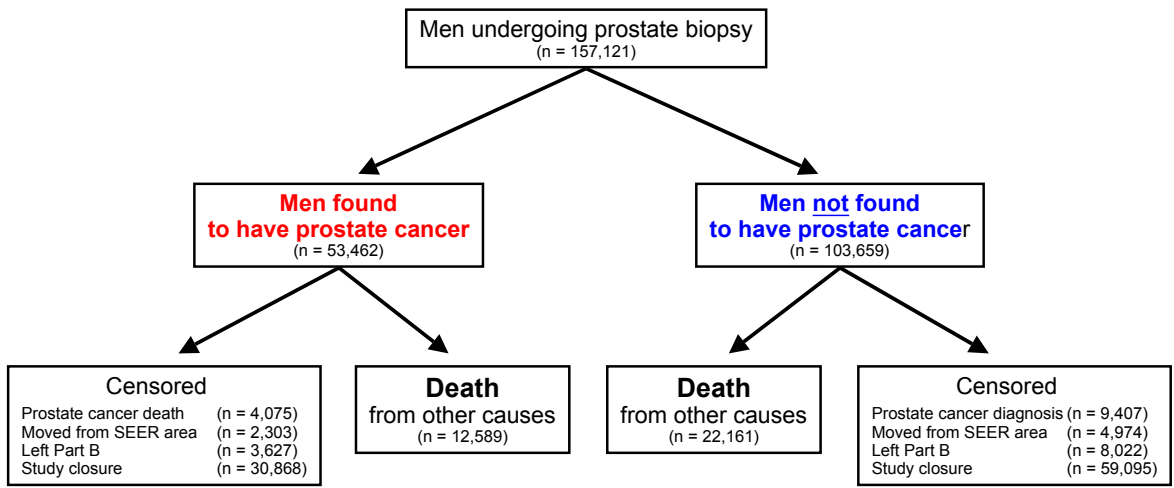

*Covariates*

The age and race of the members of our cohort was determined from the Medicare denominator file. Age is measured as the difference between the beneficiary's birth date and the date of the index biopsy.

To establish the baseline comparability between the two groups, we also measured health care utilization in the year prior to biopsy. The fact of hospital admission, the DRG weight (used to calculate hospital charge) and the length of stay were obtained from the Medicare MEDPAR (Part A) data. Inpatient and outpatient physician visits and outpatients charges were obtained from Medicare Part B. The two comorbidity measures (the Charlson comorbidity score[10] and Iezzoni's number of comorbidities[11]) were calculated based on ICD-9 diagnosis codes from both MEDPAR and Part B data. A comorbidity is identified if one condition is shown at least twice (and seven days apart) in the baseline year.

*Adjustment methods*

We compared their health care utilization in the year prior to biopsy after adjusting for expected age-race differences (men with prostate cancer would be older and more likely to be black). We used men found to have prostate cancer as the standard (referent) population and determined what utilization would have been expected for men <u>not</u> found to have prostate cancer if their age-race structure were matched to the first group. This was done with model-based direct adjustment. We used multiple linear regression (and logistic regression for the dichotomous variable of hospitalization yes/no) in which the dependent variable was the utilization variable and the independent variables were age, race and group (in which the referent category was men found to have prostate cancer). The coefficient on group was then added (or multiplied in the case of the logistic regression) to the crude data for men found to have prostate cancer to produce the adjusted data for men <u>not</u> found to have prostate cancer. This process is analogous to propensity matching.

Our primary outcome metric is the relative rate of other-cause mortality (men found vs. men <u>not</u> found to have prostate cancer). We first present the crude relative risk and then move through a series of stepwise adjustments for potential confounders. For each step men found to have prostate cancer serve as the standard population (i.e. - the crude value for this group remains

RECORDED PROSTATE CANCER MORTALITY - 10

stable throughout the adjusted analyses). This was done because we wanted to calculate the effect of transferring an excess (or deficit) in other-cause mortality to prostate cancer mortality – which, because it only occurs in the group found to have prostate cancer, cannot be adjusted.

We first adjust the relative rate simply for the expected confounders of age and race. We used two methods: a Cox Proportional Hazards model (in which age is a continuous variable) and indirect age-race adjustment (using 4 age categories and 3 race categories - producing a total of 12 age-race cells). Because these methods yielded essentially identical results, we proceeded to use the Cox model to sequentially add covariates reflecting health care utilization in the year prior to biopsy.

We also used two approaches to produce 95% confidence intervals around the relative rate: those obtained from the Cox model and by using the bootstrap method. Because these methods also yielded essentially identical results, we report those from the Cox model. Analyses were performed using either SAS 9.1 (SAS Institute Inc, Cary, NC) or STATA 10 (Stata Corp, College Station, TX).

## R<span style="font-variant:small-caps">esults</span>

**Table 1** shows the age-race structure of our study population. As expected, the average age of men found to have prostate cancer was about a year older than those <u>not</u> found to have prostate cancer (73.8 vs. 72.7 years). Also as expected, men found to have prostate cancer were slightly more likely to be black (8.6% vs. 7.1%).

**Table 1. Demographics of the study population\***

| Characteristic | Men found to have prostate cancer (N = 53,462) | Men <u>not</u> found to have prostate cancer (N = 103,659) |
|---|---|---|
| Age (mean) | 73.8 | 72.7 |
| Age Group (%) | | |
|   65-69 years | 27.2 | 33.2 |
|   70-74 years | 32.3 | 33.7 |
|   75-79 years | 23.8 | 21.3 |
|   80-84 years | 11.5 | 8.5 |
|   >85 years | 5.2 | 3.2 |
| Race (%) | | |
|   White | 82.7 | 81.1 |
|   Black | 8.6 | 7.1 |
|   Others | 8.7 | 11.8 |

\* for all comparisons $p < 0.001$

**Table 2** shows that after age-race adjustment, however, health care utilization in the year prior to biopsy was nearly identical in the two groups. Of the 8 variables measured, 2 suggested that illness burden was trivially higher in men found to have prostate cancer while 6 suggested the opposite (sign test p = .29).

Table 2. Comparison of health care utilization in the year prior to biopsy in men found and <u>not</u> found to have prostate cancer.*  All data are *age and race* adjusted.

| Characteristic[‡] | Men found to have prostate cancer [N = 53,462] (a) | Men not found to have prostate cancer [N = 103,659] (b) | Direction of effect (a – b) |
|---|---|---|---|
| Inpatient Care | | | |
| % admitted to hospital | 13.7 | 14.5 | − |
| Mean hospital charge using DRG weight ($)* | $4,868 | $4,737 | + |
| Mean LOS per admission (days) | 4.09 | 4.11 | − |
| Mean physician visits per admission | 3.00 | 3.07 | − |
| Outpatient Care | | | |
| Mean outpatient visits | 15.2 | 16.5 | − |
| Mean outpatient charges ($)† | $1,911 | $2,121 | − |
| Charlson Comorbidity Score | 0.665 | 0.632 | + |
| # of comorbidities (Iezzoni) | 0.721 | 0.735 | − |

* for all comparisons p < 0.001
† based on 2000 prices.

**Table 3** shows the crude other-cause mortality rates in the two groups and the effect of various adjustments. The age-race adjusted other-cause mortality rate was 471 per 10,000 person-years in men found to have prostate cancer vs. 468 per 10,000 in men not found to have prostate cancer (RR = 1.01; 95% CI: 0.98 - 1.03). Subsequent adjustments using the various measures of prior utilization had minimal effect.

Table 3. Other-cause mortality in men found and not found to have prostate cancer.

| Control Variables | OTHER-CAUSE MORTALITY (Rate per 10,000 person-years) | | |
|---|---|---|---|
| | Men found to have prostate cancer (N = 53,462) | Men not found to have prostate cancer (N = 103,659) | Relative Rate (95% CI) |
| Crude | 471 | 429 | 1.098 (1.07 – 1.12) |
| Age, Race | 471 | 468 | 1.006 (.98 – 1.03) |
| Age, Race, Inpatient utilization (Hospitalization, DRG weight, LOS, physician visits) | 471 | 461 | 1.021 (1.00 – 1.04) |
| Age, Race, Inpatient utilization, Outpatient utilization (visits, charges) | 471 | 458 | 1.029 (1.01 – 1.05) |
| Age, Race, Inpatient utilization, Outpatient utilization, Comorbidity scores (Charlson, Iezonni) | 471 | 461 | 1.021 (1.00 – 1.04) |

**Figure 2** shows that the other-cause survival curves for men found and <u>not</u> found to have prostate cancer are superimposed on one another – suggesting that our original hypothesis that other-cause mortality would be higher in men found to have prostate cancer was incorrect.

**Figure 2: Survival Analysis – the age-race adjusted risk of not dying from other causes in men found and <u>not</u> found to have prostate cancer.**

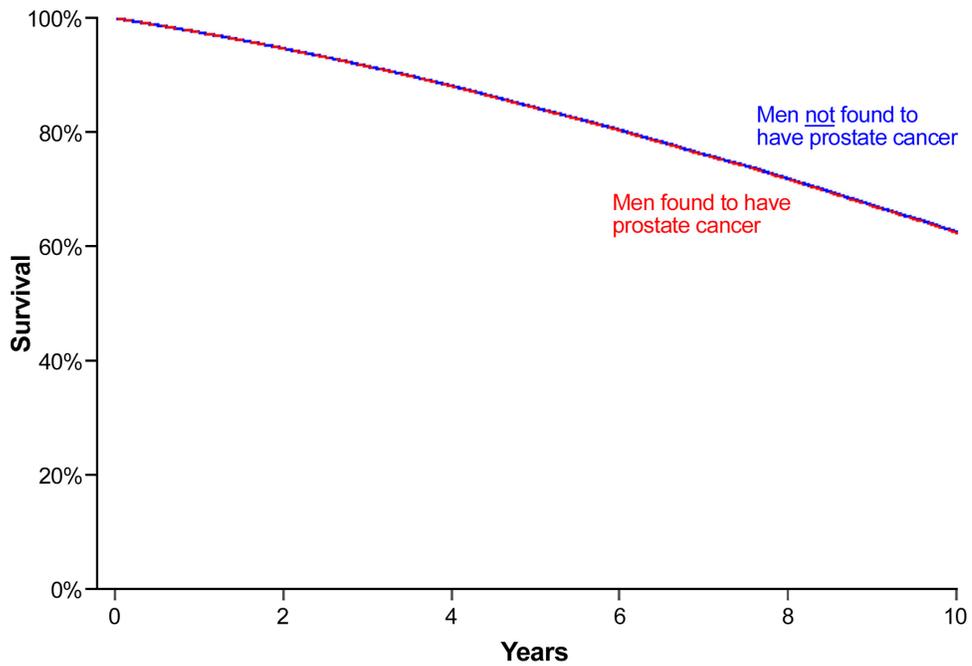

As shown in **Figure 3**, however**,** it became apparent that this effect (or lack thereof) was modified by age. In the youngest age stratum (age 65-69) there was an excess in other-cause mortality in men found to have prostate cancer, while in the oldest age stratum (age 85+) there was a deficit.

**Figure 3: Restricted Survival Analyses – the age-race adjusted risk of not dying from other causes in men found and <u>not</u> found to have prostate cancer for the two extremes of age: age 65-69 and age 85+.**

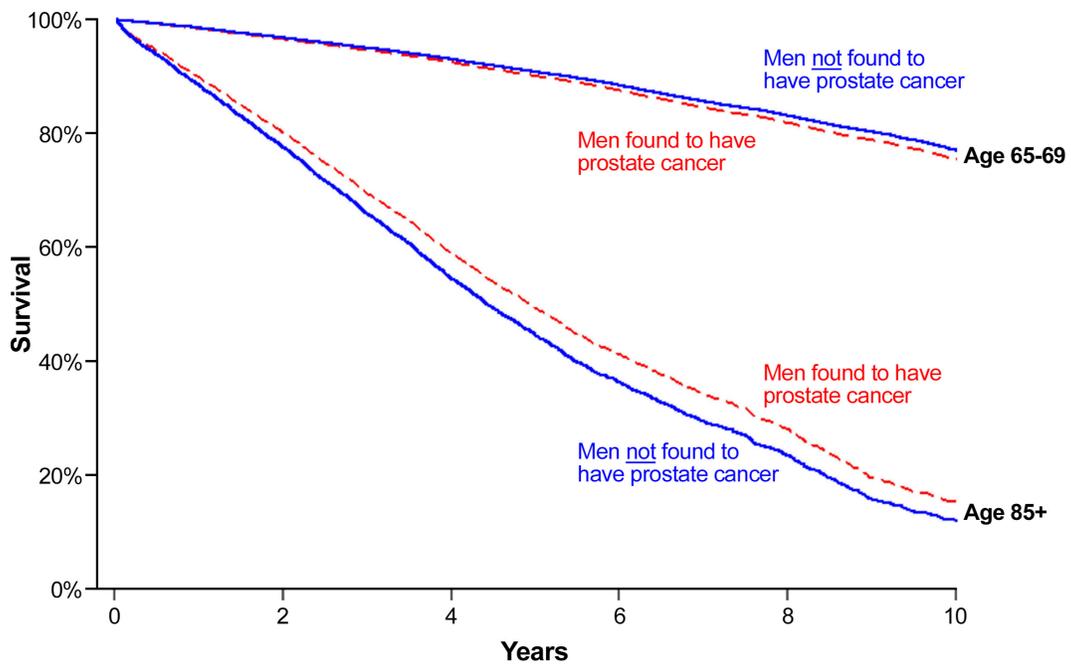

**Figure 4** shows the results of restricted analyses for each of five age strata. The relative rate of other cause mortality in men found vs. men <u>not</u> found to have prostate cancer declined in a stepwise fashion from 1.08 (95% CI 1.03 – 1.14) in men age 65-69 to 1.05 (95% CI 1.01 – 1.10) in men age 70-74 to 1.03 (95% CI 0.99 – 1.08) in men age 75-79 to 0.96 (95% CI 0.91 – 1.01) in men age 80-84 to 0.89 (95% CI 0.83 – 0.95) in men age 85 and older. The figure also shows that the proportion of men receiving definitive therapy for prostate cancer (surgery or radiation) fell dramatically with increasing age.

**Figure 4: Restricted Analyses – the age-race adjusted relative rate of other-cause mortality in men found vs. those <u>not</u> found to have prostate cancer in each of five age strata.**

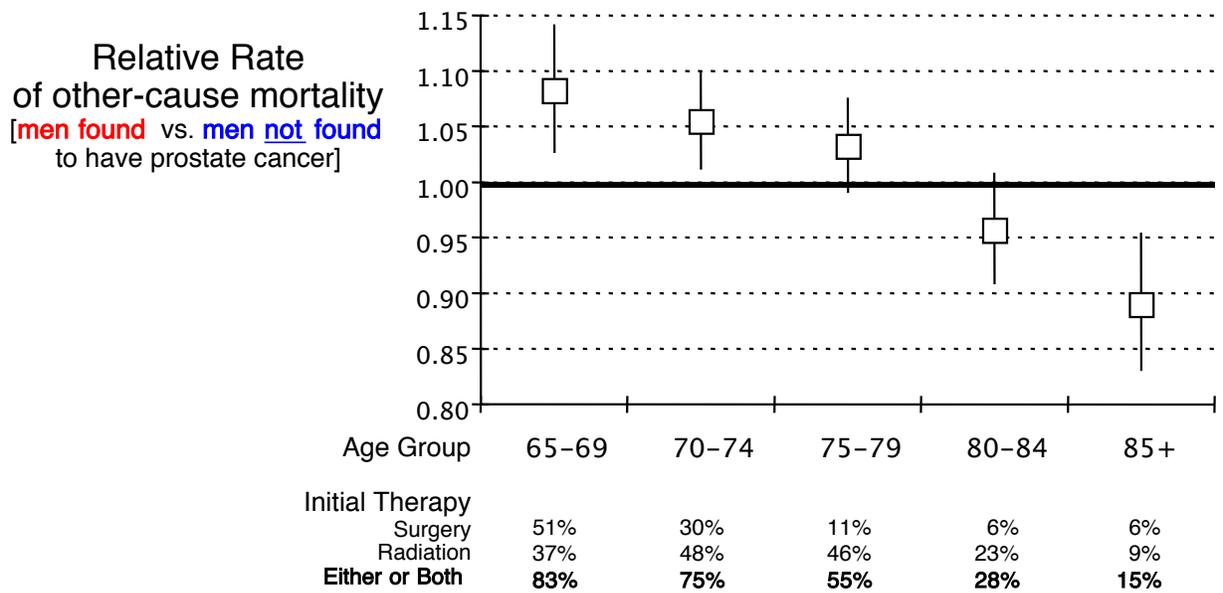

If the excess (or deficit) in other-cause mortality observed above were added to recorded prostate cancer mortality (which was 153 per 10,000 among all men found to have prostate cancer), prostate cancer mortality would rise 23% in the youngest age group (from 90 to 111 per 10,000) and would fall 30% in the oldest age group (from 551 to 388 per 10,000). The potential change in recorded prostate cancer mortality for each of the 5 age strata is shown in **Figure 5**.

**Figure 5: Percentage change in recorded prostate cancer mortality in five age strata if surplus (or deficit) in other-cause mortality were attributed to prostate cancer.**

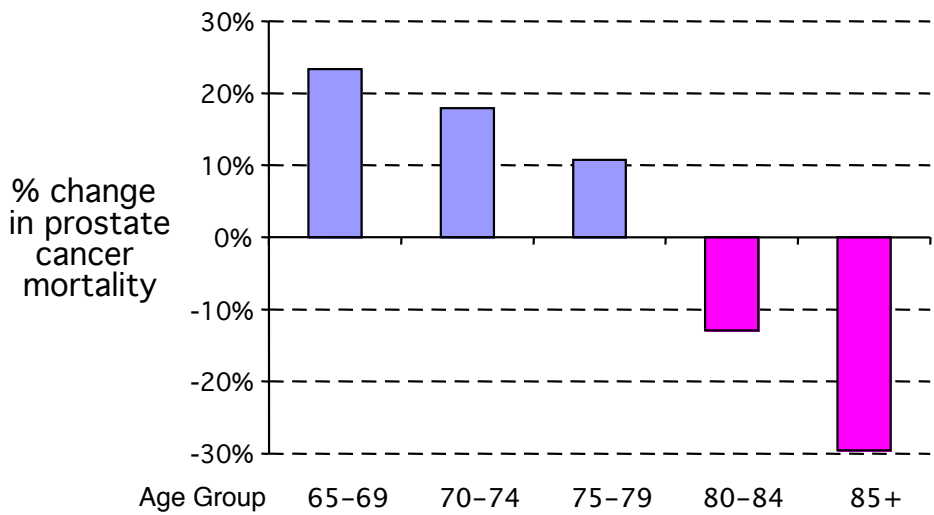

**DISCUSSION**

We investigated other-cause mortality in a cohort of over 150,000 male Medicare beneficiaries undergoing needle biopsy of the prostate during the period 1993-2001. Although we hypothesized that other-cause mortality would be higher in those found to have prostate cancer than in those who were not found to have the disease, we found no evidence to support this in our overall analysis.

We were surprised, however, that this null finding obscured a substantial effect modification by age. In younger men (age 65 to 69 and 70 to 74) other-cause mortality was significantly higher in men found to have prostate cancer. Over three-quarters of these men received definitive therapy for their disease (surgery or radiation). We suspect that the excess other-cause mortality reflects the "slippery diagnosis" bias[12] in which treatment-related deaths are attributed to other causes. We should acknowledge that, at the individual level, the cause of death may be quite ambiguous: a man dying from a heart attack in the week following a radical prostatectomy or a man dying from a pelvic abscess two years after radiation therapy to the prostate. But, at the population level, the bias becomes evident. Although the excess in other-cause mortality among younger men was relatively small (8% for men age 65-69 and 5% for men age 70-74), the potential impact on recorded prostate cancer mortality would be relatively large (an increase of 24 and 18% respectively).

In older men the effect was reversed. Among men age 85 and older other-cause mortality was significantly lower in men found to have prostate cancer. The overwhelming majority of these men did not receive definitive therapy for their disease. This deficit in other-cause mortality may represent "sticky diagnosis" bias[9] in which older men - once diagnosed - are more likely to be said to die of the disease – particularly since most are not treated. Others have suggested that the decision to treat may influence decisions about cause of death: that there may be a reluctance to attribute deaths to prostate cancer in men receiving definitive treatment and, conversely, a willingness to attribute deaths to prostate cancer (rather than something else) in men not receiving definitive treatment[13].

As with any observational study, the central concern about our findings must be confounding. Since men cannot be randomized to receive or not receive the diagnosis of cancer, inferences about the completeness of cancer mortality will necessarily depend on observational data to produce information on the mortality expected in the absence of cancer. Our approach was to examine a cohort of men who were all undergoing the diagnostic procedure used to evaluate the presence of disease and then compare other-cause mortality in those found and <u>not</u> found to have the disease. We believe this approach avoids the fundamental biases associated with a comparison of men diagnosed with prostate cancer to those in the general population (either reduced mortality due to a healthy screenee effect or increased mortality due those co-morbid conditions that may produce symptoms that lead to cancer testing). In fact, after adjusting for the expected differences found in age and race, we were able to demonstrate that the illness burden (as measured by health care utilization in the year prior to biopsy) was virtually identical. This finding held true not only for the overall analysis, but in each of the five age groups in the restricted analyses.

Our analysis has other limitations. While we have data regarding the cause of death in men diagnosed with prostate cancer (because cause of death is recorded in SEER), we do not have cause of death data in the men not diagnosed (Medicare data contains the fact of death, not the cause). Thus we cannot further enumerate the observed discrepancies in other-cause deaths within the various age groups. We also do not have data on number of needle cores obtained during the prostate biopsy – a number that undoubtedly increased over the years analyzed here. We were able to confirm, however, that our observation of effect modification by age was not sensitive to the year of diagnosis - a reasonable proxy for the increasing number of cores.

**Implications**

It is important to emphasize two inferences that should not be made based on our data. First, our data should not be taken as evidence that the observed decline in prostate cancer mortality in the United States is spurious (about 4% per year since 1994[14]). Declining mortality has been occurring within each age stratum and we have no evidence of a sufficiently large change in patterns of treatment or death attribution over time to explain these trends as being anything other than real.

Second, our data also have no bearing on the debate about whether PSA screening reduces prostate cancer mortality. Our findings are equally consistent with the possibility that PSA screening reduces prostate cancer mortality, increases it, or has no effect at all.

Instead, our findings highlight the ambiguities in determining the underlying cause of death. They reinforce the need for expansive definitions of treatment-related death by death review committees that serve those randomized trials that depend on cancer-specific mortality as a primary outcome. And that, because uncertainties are bound to persist, investigators and editors should strive to obtain a less ambiguous outcome – all-cause mortality.

**References**

[1] Extramural Committee to Assess Measures of Progress Against Cancer. Measurement of progress against cancer. J Natl Cancer Inst 1990;82:825-35.

[2] World Health Organization. Manual of the International Statistical Classification of Diseases, Injuries, and Causes of Death, based on the recommendations of the Ninth Revision Conference, 1975. Geneva: World Health Organization. 1977.

[3] INSTRUCTIONS FOR CLASSIFYING THE UNDERLYING CAUSE OF DEATH, 1995 available from the National Center for Health Statistics http://www.cdc.gov/nchs/data/2amanual.pdf

[4] Welch HG, Black WC. Are deaths within one month of cancer-directed surgery attributed to cancer? J Natl Cancer Inst 2002;94:1066-70.

[5] Brown BW, Brauner C, Minnotte MC. Noncancer deaths in white adult cancer patients. J Natl Cancer Inst 1993;85:979-87.

[6] Barry MJ, Delorenzo MA, Walker-Corkery ES, Lucas FL, Wennberg DC. The rising prevalence of androgen deprivation among older American men since the advent of prostate-specific antigen testing: a population-based cohort study. BJU Int. 2006;98:973-8.

[7] Loblaw DA, Virgo KS, Nam R, et. al. Initial hormonal management of androgen-sensitive metastatic, recurrent, or progressive prostate cancer: 2007 update of an American Society of Clinical Oncology practice guideline. J Clin Oncol. 2007;25:1596-605.

[8] Keating NL, O'Malley AJ, Smith MR. Diabetes and cardiovascular disease during androgen deprivation therapy for prostate cancer. J Clin Oncol. 2006;24:4448-56.

[9] Welch HG, Fisher ES, Gottlieb DJ, Barry MJ. Detection of Prostate Cancer via Biopsy in the Medicare – SEER Population During the PSA Era. J Natl Cancer Inst 2007;99:1395-400.

[10] Romano PS, Roos LL, Jollis JG. Adapting a clinical comorbidity index for use with ICD-9-CM administrative data: differing perspectives. J Clin Epidemiol. 1993;46:1075–1079.

[11] Iezzoni LI, Daley J, Heeren T, Foley SM, Fisher ES, Duncan C, et al.. Identifying complications of care using administrative data. Med Care. 1994;32:700–715.

[12] Black WC, Haggstrom DA, Welch HG. All-cause mortality in randomized trials of cancer screening. J Natl Cancer Inst 2002;94:167-73.

[13] Newschaffer CJ, Otani K, McDonald MK, Penberthy LT. Causes of Death in Elderly Prostate Cancer Patients and in a Comparison Nonprostate Cancer Cohort in vital statistics for prostate cancer patients. J Natl Cancer Inst 2000;92:613–21.

[14] Ries LAG, Melbert D, Krapcho M, Stinchcomb DG, Howlader N, Horner MJ, Mariotto A, Miller BA, Feuer EJ, Altekruse SF, Lewis DR, Clegg L, Eisner MP, Reichman M, Edwards BK (eds). SEER Cancer Statistics Review, 1975-2005, National Cancer Institute. Bethesda, MD, http://seer.cancer.gov/csr/1975_2005/, based on November 2007 SEER data submission, posted to the SEER web site, 2008.